\documentclass[11pt,twoside]{article} 
 
\usepackage{asp2006}
\usepackage{graphicx} 

\begin{document}
\title{The Evolution of Disk Galaxies Since z=1}  
\author{Asmus B\"ohm\altaffilmark{1,2} and Bodo L.~Ziegler\altaffilmark{2}}  
\altaffiltext{1}{Astrophysikalisches Institut Potsdam, An der Sternwarte 16,
  14482 Potsdam, Germany, email: aboehm@aip.de}
\altaffiltext{2}{Institut f\"ur Astrophysik, Friedrich-Hund-Platz 1, 37077
  G\"ottingen, Germany, email: bziegler@astro.physik.uni-goettingen.de }

\begin{abstract} 
Based on VLT/FORS spectroscopy and HST/ACS imaging, we have constructed a
sample of $\sim$\,200 field spiral galaxies that cover redshifts up to
$z \approx 1$. 
Such a large data set allows to study the evolution of fundamental galaxy parameters
like luminosity, size, mass, mass-to-light ratio etc.~as a
function of cosmic time and in various mass regimes. 
Several of our findings~--- like the
time-independent fraction of stellar-to-total mass~--- are in compliance with a hierarchical
structure growth. However, the \emph{stellar population properties} of
intermediate-redshift disks favour a
\emph{down-sizing scenario} in the sense that the average stellar ages in high-mass
spirals are older than in low-mass spirals.
\end{abstract}




\section{Introduction}   

Observations of the properties of distant galaxies at various cosmic epochs
are a powerful tool to test the predictions of cosmological simulations in the
framework of the hierarchical Cold Dark Matter paradigm.
Combining high-resolution HST/ACS imaging and deep VLT/FORS spectroscopy and
imaging, we have observed a sample of 202 disk galaxies at redshifts $0.1<z<1.0$ that
represent a mean look-back time of $\sim$\,5 Gyr. Such a data set allows - via a
comparison to local reference samples - to study the evolution of fundamental
parameters of galaxies, like luminosity, size, mass, $M/L$ ratio etc., as a function of
cosmic time. By applying models that fully account for observational effects
like seeing and the influence of the slit width, we were able to extract 
spatially resolved rotation curves from the spectra and derive the galaxies'
maximum rotation velocities as well as the total masses for 124 galaxies in
our data set. 

\section{Main Results}

In \citet{Boe04}, we reported on an earlier stage of our survey and presented
evidence for a slope change of the Tully--Fisher Relation (TFR) between 
$z \approx 0.5$ and the local universe, i.e.~a mass-dependent luminosity
evolution. Using the new, full sample of 124 galaxies, we
found that this differential evolution could be attributed to the magnitude
limit in our target
selection, but \emph{only} if the scatter of the TFR at $z \approx 0.5$ is
more than a factor of 3 larger than in the local universe 
\citep[for details, please see][]{Boe06}.

The fraction between the stellar and total mass remains roughly constant
between redshifts $z \approx 1$ and $z \approx 0$ (see left plot of Fig.~1), which
could be understood in terms of smooth accretion of dark and baryonic matter
over this epoch. If spiral galaxies already contained all their dark and
baryonic matter at $z \approx 1$, the conversion of gas into stars via continuous 
star formation would lead to an increase of the stellar mass fraction 
$M_\ast / M_{\rm vir}$ towards lower redshifts, which is not observed. 
A similar result has been found by \citet{Con05}.

The stellar mass-to-light ratios evolve more strongly for low-luminosity
spirals than for high-luminosity spirals (Fig.~1 right),
yielding evidence for an
anti-hierarchical evolution of the stellar populations (the ``down-sizing'' scenario). 
This interpretation gains further support from fits of single-zone models to
the optical and NIR colors: we find that the mean model 
stellar ages of the distant low-mass spirals are younger than those of the
distant high-mass spirals, see \citet{Fer04}. This mass-dependency of the
stellar ages
has to be counter-balanced by the evolution of other galaxy parameters to explain
the little evolution observed in the de-biased TFR slope.

\begin{figure}[t]
\vspace{0.25cm}
\includegraphics[height=6.35cm,width=5.5cm,angle=270,bb=91 54 547 612]{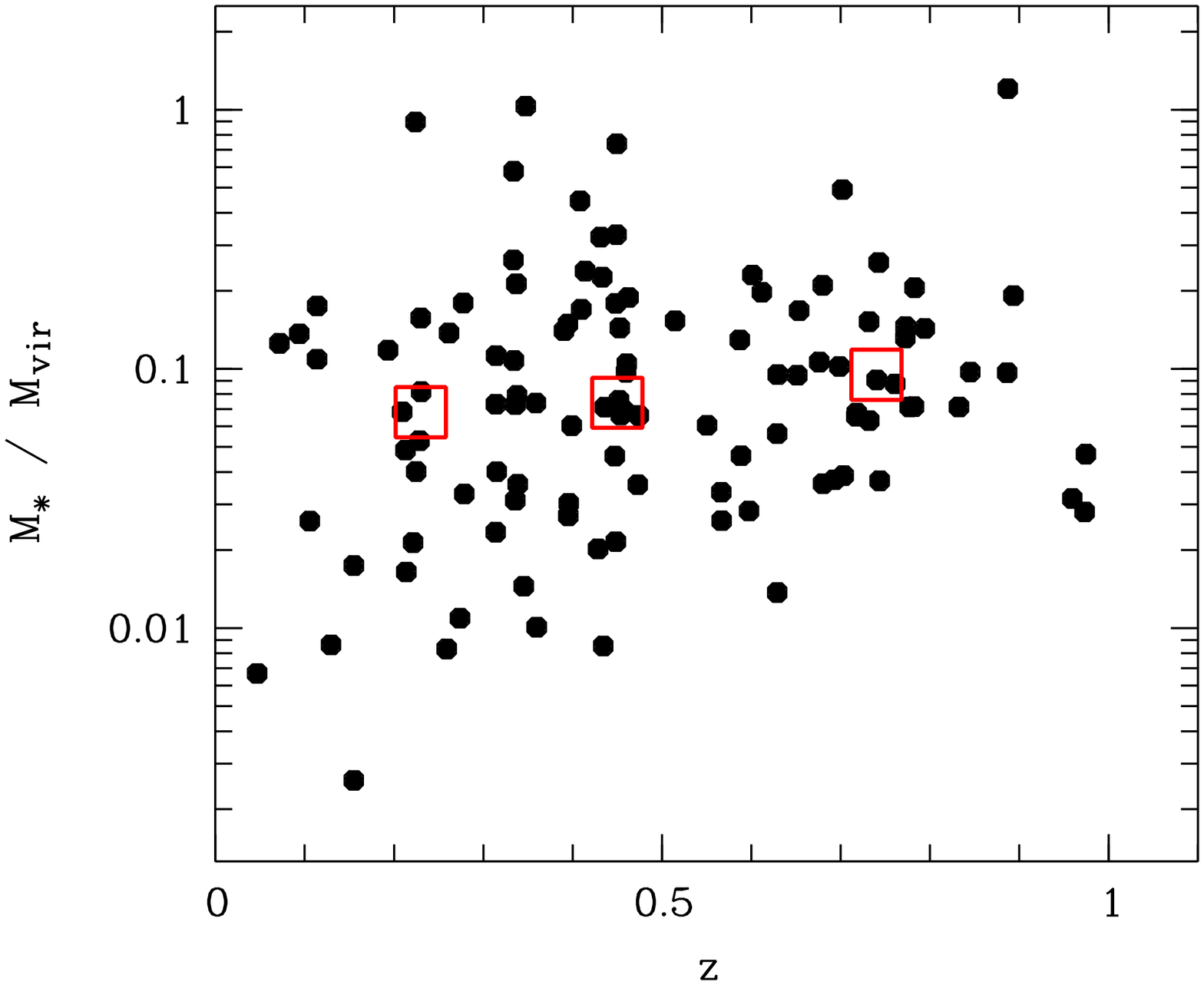}
\hfill
\includegraphics[height=6.35cm,width=5.5cm,angle=270,bb=67 41 553 622]{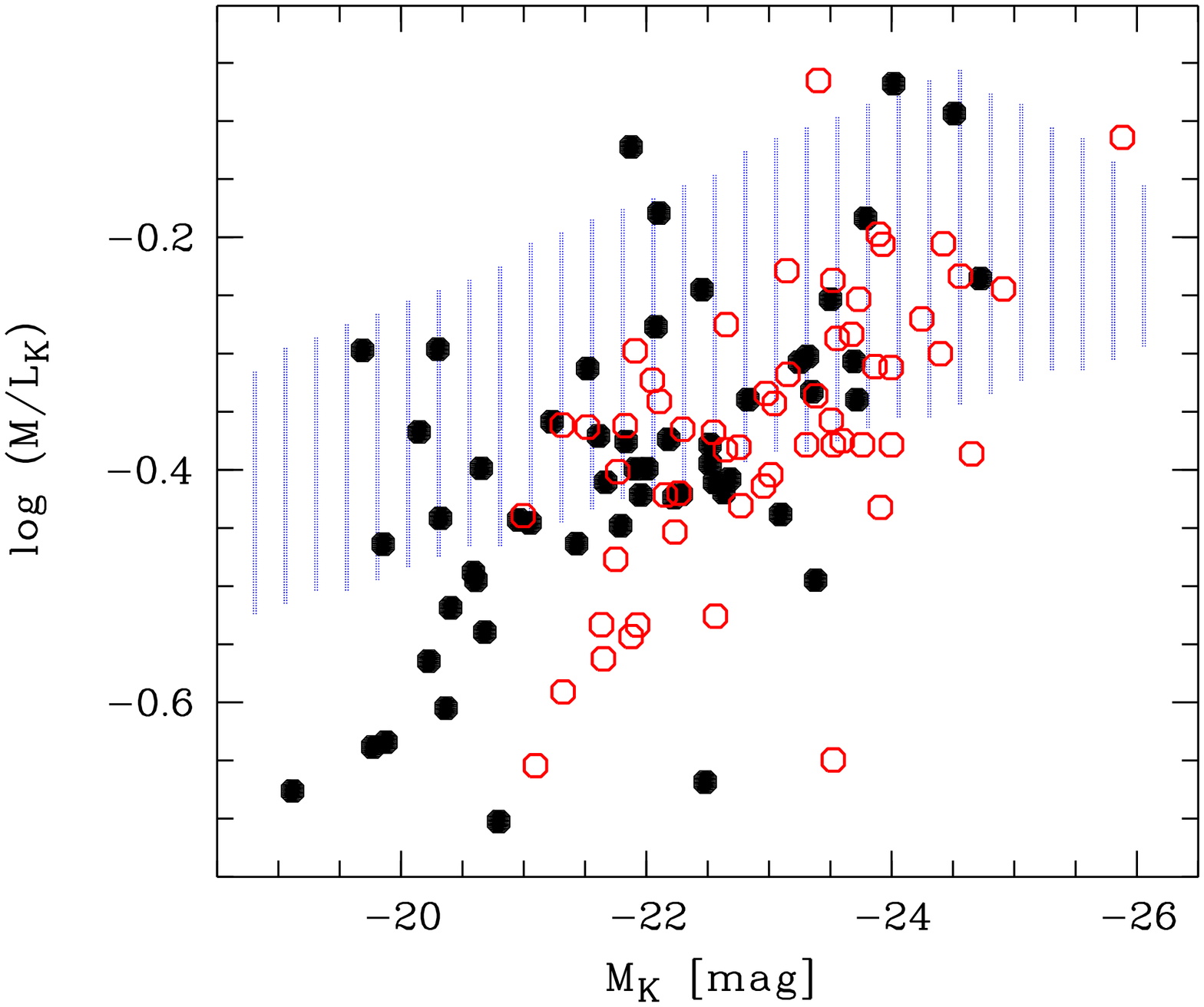}
\vspace*{0.2cm}
\caption{\emph{left:} The observed stellar mass fraction is roughly constant at
redshifts $0 < z < 1$ (squares give median values in three $z$-bins), implying the
accretion of dark (and probably baryonic) matter.
\emph{right:} Stellar mass-to-light ratios of our distant sample at
$0.1< z<0.45$ (filled circles) and $0.45<z<1.0$ (open circles), 
compared to the parameter range
covered by present-day spirals (shaded area) from \citet{Bel01}. The data
indicate a 
stronger evolution in $M/L$ for low-luminosity galaxies.
}
\end{figure}

\acknowledgements 

This work was funded by the Volkswagen Foundation (I/76\,520) and the ``Deutsches
Zentrum f\"ur Luft- und Raumfahrt'' (50\,OR\,0301).


\begin{thebibliography}{}
\bibitem[{Bell \& de Jong}(2001)]{Bel01}
{Bell, E.~F., \& de Jong, R.~S.} 2001, \apj, 550, 212

\bibitem[{B\"ohm} {et~al.}(2004)]{Boe04}
B\"ohm, A., Ziegler, B.~L., Saglia, R.~P., et al. 2004, \aap, 420, 97

\bibitem[{B\"ohm \& Ziegler}(2006)]{Boe06}
{B\"ohm, A., \& Ziegler, B.~L.} 2006, \apj, submitted, (astro-ph/0601505)

\bibitem[{Conselice} {et~al.}(2005)]{Con05}
Conselice, C.~J., Bundy, K.~E., Richard, S., et al.~2005, \apj, 628, 160

\bibitem[{Ferreras} {et~al.}(2004)]{Fer04}
{Ferreras, I., Silk, J., B\"ohm, A., \& Ziegler, B.~L.} 2004, \mnras, 355, 64

\end{thebibliography}
\end{document}